\documentclass[10pt]{article}

\usepackage{amsmath}
\usepackage{amssymb}

\usepackage{graphicx}

\usepackage{cite}

\usepackage{color} 

\usepackage[labelfont=bf,labelsep=period,justification=raggedright]{caption}

\usepackage{epsfig}
\usepackage{amscd}
\usepackage{amsmath}
\usepackage{amssymb}
\usepackage{subfigure}

\newlength{\arrayrulewidthOriginal}
\newcommand{\Cline}[2]{%
  \noalign{\global\setlength{\arrayrulewidthOriginal}{\arrayrulewidth}}%
  \noalign{\global\setlength{\arrayrulewidth}{#1}}\cline{#2}%
  \noalign{\global\setlength{\arrayrulewidth}{\arrayrulewidthOriginal}}}

\makeatletter
\renewcommand{\@biblabel}[1]{\quad#1.}
\makeatother

\date{}

\begin{document}

\begin{flushleft}
{\Large
\textbf{Population physiology: leveraging population scale (EHR) data to
  understand human endocrine dynamics}
}
\\
D. J. Albers$^{1,\ast}$, George Hripcsak$^{2}$, Michael Schmidt$^{3}$
\\
\bf{1} Department of Biomedical Informatics, Columbia University,
$622$ W $168^{th}$ St. VC-5, New York, NY 10032
\\
\bf{2} Department of Biomedical Informatics, Columbia University,
$622$ W $168^{th}$ St. VC-5, New York, NY 10032
\\
\bf{3} Department of Neurology, Columbia University, 177 Fort
Washington, 8GS-331, New York, NY 10032
\\
$\ast$ E-mail: Corresponding david.albers@dbmi.columbia.edu
\end{flushleft}

\section*{Abstract}
Studying physiology over a broad population for long periods of
time is difficult primarily because collecting human physiologic data
is intrusive, dangerous, and expensive. One solution is to use data
that has been collected for a different purpose. Electronic
health record (EHR) data promise to support the development and
testing of mechanistic physiologic models on diverse populations and
allow correlation with clinical outcomes, but limitations in
the data have thus far thwarted such use. For example, using
uncontrolled population-scale EHR data to verify the outcome of time
dependent behavior of mechanistic, constructive models can be
difficult because: (i) aggregation of the population can obscure or
generate a signal, (ii) there is often no control population with a
well understood health state, and (iii) diversity in how the
population is measured can make the data difficult to fit into
conventional analysis techniques. This paper shows that it is possible
to use EHR data to test a physiological model for a population and
over long time scales. Specifically, a methodology is developed and
demonstrated for testing a mechanistic, time-dependent, physiological
model of serum glucose dynamics with uncontrolled, population-scale,
physiological patient data extracted from an EHR repository. It is
shown that there is \emph{no observable} daily variation the
normalized mean glucose for any EHR subpopulations. In contrast, a
derived value, daily variation in \emph{nonlinear correlation}
quantified by the time-delayed mutual information (TDMI), \emph{did
  reveal} the intuitively expected diurnal variation in glucose levels
amongst a wild population of humans. Moreover, in a population of
intravenously fed patients, there was no observable TDMI-based diurnal
signal. These TDMI-based signals, via a glucose insulin model, were
then connected with human feeding patterns. In particular, a
constructive physiological model was shown to correctly predict the
difference between the general uncontrolled population and a
subpopulation whose feeding was controlled.


\textbf{keywords:} time-delay mutual information | nonlinear
time-series analysis | information theory | high dimensional data |
electronic health record | time-delay dynamical systems | physiology |
non-uniform sampling



\section*{Introduction}

Human physiology, as a science, aims to understand the mechanical,
physical, and biochemical functions of humans; moreover, because human
dynamics transpire both on multiple spatial scales, ranging from
molecular (e.g., genetics), to cell (e.g., metabolism), to organ
(e.g., the heart \cite{peskin_heart_simulate}), to collections of
organs (e.g., the circulatory system) and on multiple time scales
ranging from fractions of a second to decades, it is likely that
complete models of human functioning will consist of highly complex
models whose scales interact in complex ways (e.g., via nonlinear
resonance) \cite{levin_complex_systems_bio_ams}. In this context,
\emph{population physiology} aims to understand medium to long time
scales of human physiology where \emph{a population of humans} is
required to construct or discover a signal\footnote{Metaphorically,
  population physiology is to physiology as climatology is to
  weather.}. Moreover, once a signal is constructed the goal is to use
this signal understand human dynamics by both understanding the
sources of the signals and then stratifying the population into
meaningful classes (e.g., phenotypes) according to the different
signals. Consequently, population physiology, as we
conceive it, has two broad features: data analysis consisting of the
construction and analysis of population scale physiological signals,
and the mechanistic modeling that can explain and rationalize those
signals.

The mathematical modeling of physiological systems on the cellular and
organ scales has a long history (c.f., \cite{keenerI} and
\cite{keenerII} for a wonderful introduction), while the modeling
larger scale organ structures is just beginning
\cite{lncc_circulatory}. Fundamental to mathematical modeling of
physiology is a concrete connection to real data; as is the case for
other basic sciences, mathematical physiological modeling is always
tested against physiological data collected in rigorously controlled
circumstances. Nevertheless, there are at least two elements missing
from modern physiological analysis, analysis over large populations,
and analysis over long time periods. The former is important because
human beings have diverse reactions to different inputs (e.g., drugs,
foods, etc), and those differences have their roots in physiology. The
later is important because many differences amongst human reactions to
input occur on a slow time-scale; for instance, some smokers develop
cancer while others do not. And again, these differences can have
their roots in physiology. The problem with using the classical
physiology framework with its rigorously controlled conditions to
study a large population over a long time period is that it is too
expensive, intrusive, and dangerous to collect physiologic data for a
large population over a long time period. Thus, it is likely that the
lack of availability of population scale, long term data is the
primary reason why wide-population, long term, physiologic studies to
not exist.

With the advancement of electronic health records (EHR) repositories,
the ``lack of data'' problem will be replaced with data analysis and
data mining problems. Electronic health records hold data for large,
diverse populations, and they cover periods of decades
\cite{ehr_use_1} \cite{ehr_use_2} \cite{ehr_fallacies}. Nevertheless,
despite years of work, the methods needed to exploit EHR data remain
in their infancy. A necessary realization for using EHR data is
recognizing that the EHR represents a natural system in its own
right. In particular, EHR data not only represents the physiology of
the diverse population being cared for, but also the following: the
hospital measurement dynamics (e.g., individual hospital protocols);
the local environment (e.g., exposure to pollutants); local customs
(e.g., willingness to seek medical attention); and any other features
of the environment in which the data is collected.  To see some of the
difficulties and potential associated with the analysis of EHR data,
consider four notably relevant examples: (i) Ref. \cite{warfarin_ehr}
demonstrates the limitations of using general population EHR data for
estimating drug dosages; (ii)
Ref. \cite{george_ehr_classification_limitations} reveals difficulties
with using general EHR data for classification of disease (i.e.,
community-acquired pneumonia) often required manual intervention to
achieve accurate results; Ref. \cite{ehr_fallacies} outlines various
factors that will constrain EHR data; and
Ref. \cite{pnas_red_blood_cells_anemia} demonstrates that relevant,
predictive, phenomenological master equations of physiological
functioning (concentrations of red blood cells) can be generated using
data that \emph{could} exist in an EHR repository (note that in
Ref. \cite{pnas_red_blood_cells_anemia} the population dynamics refers
to a population of red blood cells) and that, if integrated into a EHR
infrastructure, would help with early prevention of disease (i.e.,
anemia). Advancing such methods is a stepwise process, and here we
present what we believe is an important early step: showing that it is
feasible to use EHR data in conjunction with a constructive
physiological model --- specifically, that we can test a physiologic
model with an EHR data-derived signal.

To study how EHR data can be used in conjunction with a physiological
model, we consider the relatively simple problem of glucose variation
because it is easy to present and understand, it has relevant, well
understood models, and because we know what the answer should
be. Specifically, we leverage the following tools or data sets: (i) a
subpopulation patients with at least two glucose measurements from an
EHR that includes all in-patients and outpatients seen at an academic
medical center over 20 years; (ii) a well sampled patient from the
same previously mentioned EHR; (iii) a set of particularly sick,
continuously fed, immobile, comatose patients taken from the neural
intensive care unit (NICU) portion of the previously mentioned EHR;
(iv) a relatively simple mechanistic glucose-insulin model with
various different feeding regiments; and (v), the time-delay mutual
information (TDMI) which quantifies \emph{nonlinear correlation}
between \emph{ensembles} of measurements separated by a given amount
of time.

Along with demonstrating that EHR data can be used to test physiologic
models for populations over long time periods, we also discover that
while human glucose levels are highly aperiodic, there is nevertheless
a stable, long term diurnal structure in the \emph{nonlinear
  correlation} between glucose values separated in time in healthy,
``wild'' humans. Moreover, while it is likely that many features
contribute to the observed diurnal cycle in correlated glucose, only
\emph{two} interacting time-scales are required to reproduce the
observed diurnal signal --- a ``statistically periodic'' feeding
regiment that exists on the scale of weeks and the organ level
dynamics that exists on the order of minutes. Less broadly, we find
that: (i) to first order statistical moment (e.g., the mean), daily
variation in the TDMI is a function of feeding alone---no diversity in
other parameters that determine glucose/insulin regulation are
required;(ii) that glucose regulation acts like a control system on a
fast time scale (order of minutes) in contrast to kidney function
which behaves like a filtering system
\cite{stat_dyn_diurnal_correlation_short}; (iii) a diurnal signal in a
derived value, nonlinear correlation (TDMI), that can be used to distinguish different populations; and
(iv) it is possible to circumvent inter-patient variability though
aggregating populations, but one must be very careful interpreting the
results \cite{patient_aggregation_paper}.

\section*{Materials and Methods}

\subsection{Ethics statement}
This work was approved by the Columbia University Institutional Review
Board. Informed consent was waived by the Institutional Review Board for
this retrospective research.

\subsection{Glucose-Insulin physiology}
There are many ways to conceptualize the portion of the endocrine
system that controls glucose/insulin regulation; for our purposes we will
consider two, the more concrete picture of the endocrine system as a
mechanistic, cellular, physiological machine and the more abstract
picture of the endocrine system as a control system
\cite{feedback_control_intro} for glucose/insulin regulation.

Beginning with the concrete mechanistic model, we use the model
presented in Ref. \cite{sturis_91} which consists of six ODEs,
specifically:
\begin{align}
\frac{dI_p}{dt}& = f_1(G) - E(\frac{I_p}{V_p} - \frac{I_i}{V_i}) -
\frac{I_p}{t_p} \\
\frac{dI_i}{dt} &= E(\frac{I_p}{V_p} - \frac{I_i}{V_i}) -
\frac{I_i}{t_i} \\
\frac{dG}{dt} &= f_4(h_3) + I_G(t) - f_2(G) -f_3(I_i)G 
\end{align}
and a three stage linear filter:
\begin{align}
\frac{dh_1}{dt} &= \frac{I_p - h_1}{t_d} \\
\frac{dh_2}{dt} &= \frac{h_1 - h_2}{t_d} \\
\frac{dh_3}{dt} &= \frac{h_2-h_3}{t_d}
\end{align}
where the state variables correspond to: $I_p$, plasma insulin; $I_i$,
remote insulin; $G$, glucose; and $h_1$, $h_2$ and $h_3$ which
correspond to three parameterized delay processes. The \emph{major}
parameters include: (i) $E$, a rate constant for exchange of insulin
between the plasma and remote compartments; (ii) $I_G$, the exogenous
(externally driven) glucose delivery rate; $t_p$, the time constant
for plasma insulin degradation; (iii) $t_i$, the time constant for the
remote insulin degradation; (iv) $t_d$, the delay time between plasma
insulin and glucose production; (v) $V_p$, the volume of insulin
distribution in the plasma; (vi) $V_i$, the volume of the remote insulin
compartment; (vii) $V_g$, the volume of the glucose space; (viii)
$f_1(G)=\frac{R_m}{1- \exp(\frac{-G}{V_g c_1} + a_1)}$, insulin
secretion; (ix) $f_2(G)= U_b(1- \exp(\frac{-G}{C_2V_g}))$,
insulin-independent glucose utilization; (x) $f_3(I_i)=\frac{1}{C_3
  V_g}( U_0 + \frac{U_m - U_0}{1 + (\kappa I_i)^{-\beta}})$,
insulin-dependent glucose utilization ($\kappa = \frac{1}{C_4}
(\frac{1}{V_i} - \frac{1}{E t_i})$); and (xi) $f_4(h_3)=\frac{R_g}{1 +
  \exp(\alpha (\frac{h_3}{C_5 V_p} -1))}$, insulin-dependent glucose
utilization. Note that a full list of the parameters in this model, as
well the model parameter settings used in this paper, can be found in
table \ref{table:model_parameters}; moreover, Ref. \cite{keenerII}
provides a nice discussion of this particular model. With the
exception of the exogenous glucose delivery rates, which we will
discuss shortly, we utilize all the standard parameter settings used
in Ref. \cite{sturis_91}. Finally, there do exist more complex, higher
order glucose/insulin metabolism
models\cite{review_glucose_models_06}, but because the point was to
choose the simplest system of ODEs that can be used to represent the
data-driven signal, we chose this rather standard model.

Reaching beyond this model, it is important to note that a complete
physiological understanding of the endocrine system, or even the
glucose/insulin cycle, has not yet been achieved. For instance, how
insulin reacts at the plasma membrane of insulin sensitive cells is
still poorly understood (for other examples, c.f.,
\cite{glucose_paradox_84} \cite{vale_pnas_2009}). With respect to
diurnal cycles in glucose/insulin dynamics, the following effects have
been observed: in \emph{fasting humans}, there are wake-sleep cycle
based effects on pancreatic enzyme secretions
\cite{circ_fasting_humans}; physical activity has an effect on insulin
secretion \cite{circ_mobility_humans}; and in rats there appears to be
an endogenous circadian oscillator (internal clock) located within the
pancreatic islets \cite{rat_circ_insulin}. Most importantly, it is
well understood that nutrition intake is the primary first order
driver of the glucose-insulin cycle \cite{circ_mobility_humans} (hence
the need to use fasting humans as a control to isolate the more
sensitive glucose-insulin effects.) All of these studies were carried
out under the classical physiology framework. Moreover, to resolve
many of the previously listed signals required rigorous control of the
measured individuals---most EHR data will never meet these
standards. But, the noted contrast between classical physiology data
and EHR data helps clarify the goal of this paper: we are not trying
to discover an ultra-sensitive, controlled, physiological effect that
is resolvable over a short time period; rather, we are trying to
discover what can be resolved with EHR data. Specifically, we are
trying to discover gross, long term, population-wide effects that have
the potential to help stratify populations into observably different
types --- types that can eventually be linked to different health
states. Moreover, because the individuals within the EHR have
observably differing health states that do not require ultra-fine
resolution to observe, the hope is that we will be able to eventually
use EHR data to discover and categorize different, long term,
physiologic macrostates. Hence the justification for not choosing the
most complicated glucose/insulin model. While the model we utilize
parameterizes away many of these higher-order effects, it remains
driven by nutrition, the source of the first order, elementary
glucose/insulin dynamics we are trying to verify.


To interpret the results, it will help to abstract the physical
mechanisms to a control system. In particular, the regulation of
glucose can be thought of as an intra-body feedback control system
where the body \emph{has a goal of maintaining a constant
  concentration of glucose} and attempts to achieve this goal via
various physiological mechanisms. Broadly, when glucose levels are
high, insulin is released by the pancreas leading to glucose being
stored in the liver faster than it is released \emph{and} the rate at
which glucose is metabolized by the body is increased. Similarly, when
glucose levels are low, glucagon is released by the pancreas, allowing
for an increase in the rate glucose is released from the liver as well
as a decrease in the rate glucose is metabolized by the body. This
contrasts with, for example, the kidneys and their relation with
creatinine, which can be grossly thought of as a filtering system
instead of a control system aiming at maintaining a particular level
of glucose.

The only part of the model we vary is the external driving, or the
\emph{exogenous glucose delivery rate}, $I_G(t)$; specifically, we
consider four different feeding regiments. The first feeding regiment
we consider is a population that is feed continuously and where each
member of the population is fed at a different rate. This feeding
regiment is meant to simulate an intensive care unit population and is
denoted by the feeding function $I_{G,cp}$. The other three feeding
regiments are based on simulated meals. To construct mealtime feeding
structure, begin by defining the set of meal times, specified by the
set $M = \{ m_1, \cdots, m_n \}$, where the $m_i$'s represent times
over a $24$-hour interval, and $n$ is the number of meal times within
a $24$-hour period.  Next define the exogenous glucose delivery rate
at the current time, $t_c$, as:
\begin{equation}
I_G(t<t_c) = \sum_i^N I_j e^{\frac{k}{t-m_i}}
\end{equation}
where $I_j$ is the peak rate of delivery of glucose for a given
individual $j$ at time $m_i$, $N=\#\{m_i<t_c\}$ represents the total
number of meals that have passed by time $t$, and $k$ is the decay
constant ($k=0.5$). The decay constant is set such that the meal
persists over about two hours, a time that is considered realistic
\cite{sturis_91}. Next, relative to the $m_1 = 8$, $m_2=12$, and
$m_3=18$, define the following three feeding regiments: \emph{periodic
  individual}, $M_{pi} = [m_1, m_2, m_3]$; \emph{noisy individual},
$M_{ni} = [m_1 + \nu_1(k), m_2 + \nu_2(k), m_3 + \nu_3(k)]$ where
$\nu_i(k)$ is a uniform random variable on the interval $[-1,1]$ and
$k$ represents an integer day (implying that $\nu_i$ changes every
day); and \emph{random individual}, $M_{ri} = [\nu_1(k), \nu_2(k),
\nu_3(k)]$ where $\nu_i(k)$ is a random (non-repeated) integer on the
interval $[0, 23]$ and $k$ is again an integer day (implying that
$\nu_i$ changes every day). We now have four feeding regiments,
\emph{continuously fed population} ($I_{G, cp}$), a periodically fed
individual ($I_{G,pi}$), a noisy-periodic individual ($I_{G, npi}$),
and a random individual ($I_{G, ri}$), defined formally as:
\begin{align}
I_{G, cp} &= I_j \text{ constant } \in [100,225] \text{ mg/min} \\
I_{G, pi}(t) &= \sum_i^N I e^{\frac{k}{t-m_i}}, I=216 \text{ mg/min,}
m_i \in M_{pi} \\
I_{G, ni}(t) &= \sum_i^N I e^{\frac{k}{t-m_i}}, I=216 \text{ mg/min,}
m_i \in M_{ni} \\
I_{G, ri}(t) &= \sum_i^N I e^{\frac{k}{t-m_i}}, I=216 \text{ mg/min,}
m_i \in M_{ri} 
\end{align}
These four different driving mechanisms reflect what we believe to be
a relatively minimalistic amount of variation within the
glucose/insulin model parameter and function space.

\renewcommand{\arraystretch}{1.5}
\begin{center}
\small
\begin{table}[!ht]
\begin{tabular}{|l|l|}
\hline
\multicolumn{2}{|c|}{\textbf{Glucose model parameters}} \\
\Cline{2pt}{1-2} \hline
\Cline{2pt}{1-2}
$V_p$  & $3$ l  \\ \hline \hline
$V_i$  & $11$ l  \\ \hline \hline
$V_g$ & $10$ l  \\ \hline \hline
$E$  & $0.2$ l min$^{-1}$ \\ \hline \hline
$t_p$  & $6$ min  \\ \hline \hline
$t_i$  & $100$ min \\ \hline \hline
$t_d$  & $12$ min  \\ \hline \hline
$R_m$  & $209$ mU min$^{-1}$  \\ \hline \hline
$a_1$  & $6.67$ \\ \hline \hline
$C_1$  & $300$ mg l$^{-1}$\\ \hline \hline
$C_2$  & $144$ mg l$^{-1}$  \\ \hline \hline
$C_3$  & $100$ mg l$^{-1}$  \\ \hline \hline
$C_4$  & $80$ mU l$^{-1}$ \\ \hline \hline
$C_5$  & $26$ mU l$^{-1}$  \\ \hline \hline
$U_b$  & $72$ mg min$^{-1}$  \\ \hline \hline
$U_0$  & $4$ mg min$^{-1}$ \\ \hline \hline
$U_m$  & $94$ mg min$^{-1}$  \\ \hline \hline
$R_g$  & $180$ mg min$^{-1}$  \\ \hline \hline
$\alpha$  & $7.5$ \\ \hline \hline
$\beta$  & $1.77$ \\ \hline \hline
\end{tabular}
\caption{Full list of parameters for the glucose/insulin model
  \cite{sturis_91} used in this paper; note that these are the model
  parameters we us in this paper.}
\label{table:model_parameters}
\end{table}
\end{center}

\subsection{Data composition}

We consider the time series of glucose measurements of two real
populations of humans extracted from the Columbia University Medical
Center (CUMC) EHR: (i) the time series of glucose measurements
extracted from an EHR for all inpatients and outpatients over 20 years
($800,000$ patients with roughly $12,000,000$ glucose measurements);
(ii) the time series of glucose measurements for a small subset of
patients ($43$ in total) seen in the NICU who are continuously fed,
immobile, and comatose---note that this cohort of patients is
represented by between $4$ and $193$ measurements taken on the order
of minutes to hours (many patients have approximately a weeks' worth
of hourly measurements). There are four important differences in these
populations: (i) population one is uncontrolled and monitored poorly
(it is the general patient population after all) whereas population
two is highly controlled and monitored; (ii) population one has an
unknown and uncontrolled feeding regiment whereas population two is
being fed continuously and in a very controlled and documented
fashion; (iii) population one represents a diverse set of humans with
diverse and unknown health states whereas population two represents a
very sick population whose degree of acuity is considerably higher and
more narrowly defined than that of population one; and (iv) while the
detailed understanding of metabolic function is unknown in both
populations, it is very likely that the metabolic functioning of
patients in population two is substantially more compromised. Thus,
population two functions roughly as a control to isolate the effects
of continuous feeding on glucose daily variability because this
population has relatively few normal external physiological forcing
mechanisms (e.g., sleep cycle, daily exercise, etc). In contrast,
population one is meant to represent the population at large whose
feeding regiment is uncontrolled, highly discontinuous, and has
unknown regularity.

In addition to the two populations, we have included two relatively
densely and uniformly measured, non-ICU patient from the CUMC EHR. We
have included these patients to demonstrate that, despite
population-aggregation effects on glucose variability (recall that
Ref. \cite{stat_dyn_diurnal_correlation_short} detailed how
aggregation of different sources can affect a TDMI signal) the diurnal
TDMI peaks do occur in regular patients. These patients were selected
from among the $100$ patients with the most glucose values in the
CUMC EHR and they represent the two typical types of patients ---
patients with high and patients with low diurnal TDMI peaks. Neither
patients glucose measurements come primarily from the ICU setting. We
\emph{speculate} that the magnitude of the diurnal peak is related to
the relative health of the pancreas (or successful glucose regulation)
because the patient with the relatively small diurnal TDMI-based peak
has a failing pancreas. Nevertheless, due to the complexity of the
models and patients, resolving the source of the higher order features
of the TDMI distribution (e.g., the higher order moments) of the
$24$-hour TDMI peaks among patients is beyond the scope of paper.

\subsection{Computational methods}

We use two diagnostics for the EHR and model glucose time series, (i)
intra-patient normalized glucose by hour, and (ii) the TDMI of the
glucose time series (Ref. \cite{patient_aggregation_paper} explains
how the TDMI can be applied to a population).

With respect to (i), we normalize each patient to mean zero and unit
variance, and then calculate the mean and variance of glucose by hour
over the population.  We do this because there is a high degree of
individual variability within each population, and individuals were
measured differently from each other. Therefore, to resolve a property
such as the by-hour daily variation of glucose values, we must remove
inter-individual variation. Without this correction, inter-individual
variation and therefore population aggregation effects became the
first order effects.

With respect to (ii), we calculate the TDMI \cite{kantz_book},
\cite{sprott_book}, given by:
\begin{equation}
I(x_t, x_{t-\delta t}) = - \int p(x_t, x_{t-\delta t}) \log
\frac{p(x_t, x_{t-\delta t})}{p(x_t)p(x_{t-\delta t})}dx_t
dx_{t-\delta t}
\end{equation}
where $x_t$ and $x_{t-\delta t}$ represent an ensemble of \emph{all the
  intra-patient pairs of points in the population of time series
  separated by a time $\delta t$} and $p(\cdot)$ denotes the PDF of
those ensembles; note that the TDMI captures linear and nonlinear
correlations in time, which differs from, say, auto or linear
correlation calculations (to see this applied to kidney function, see
\cite{stat_dyn_diurnal_correlation_short}, and for general
application, see \cite{patient_aggregation_paper}).  Finally, to
calculate the TDMI, one must estimate the joint and marginal PDFs,
here we used a KDE estimation routine \cite{kde_matlab_I} implemented
on MATLAB.

In general, the TDMI is a unit-less quantity; a TDMI of $0$ (within
bias) implies that there is no correlation between sequential values
in a time series for a given $\delta t$. TDMI values begin to become
important when they exceed the expected bias associated with
calculating the mutual information, which is approximately
$\frac{1}{M}$ where $M$ is the number of pairs of points used to
estimate the TDMI ($\sim 0.001$ in this experiment).  With a perfect
correlation between sequential values, the TDMI will be equal to the
entropy (or auto-information) of the series, which is numerically equal
to the TDMI at $\delta t = 0$ (and is calculated automatically as part
of the experiment). In this experiment the entropy was about $0.85$
and represented the maximum TDMI. (In most of our experiments, the
entropy is in the $0.5$ to $2$ range.)

With respect to the models, the ODEs were integrated over time-periods
ranging from seven days to three weeks.  A standard fourth-order
Runga-Kutta integration routine, with a step-size of $10^{-4}$, was
utilized.

\section*{Results}

\subsection{Basic physiological synopsis}

\begin{figure}[!ht]
  \subfigure[Glucose-insulin model with continuous feeding and glucose
  response]{
    \epsfig{file=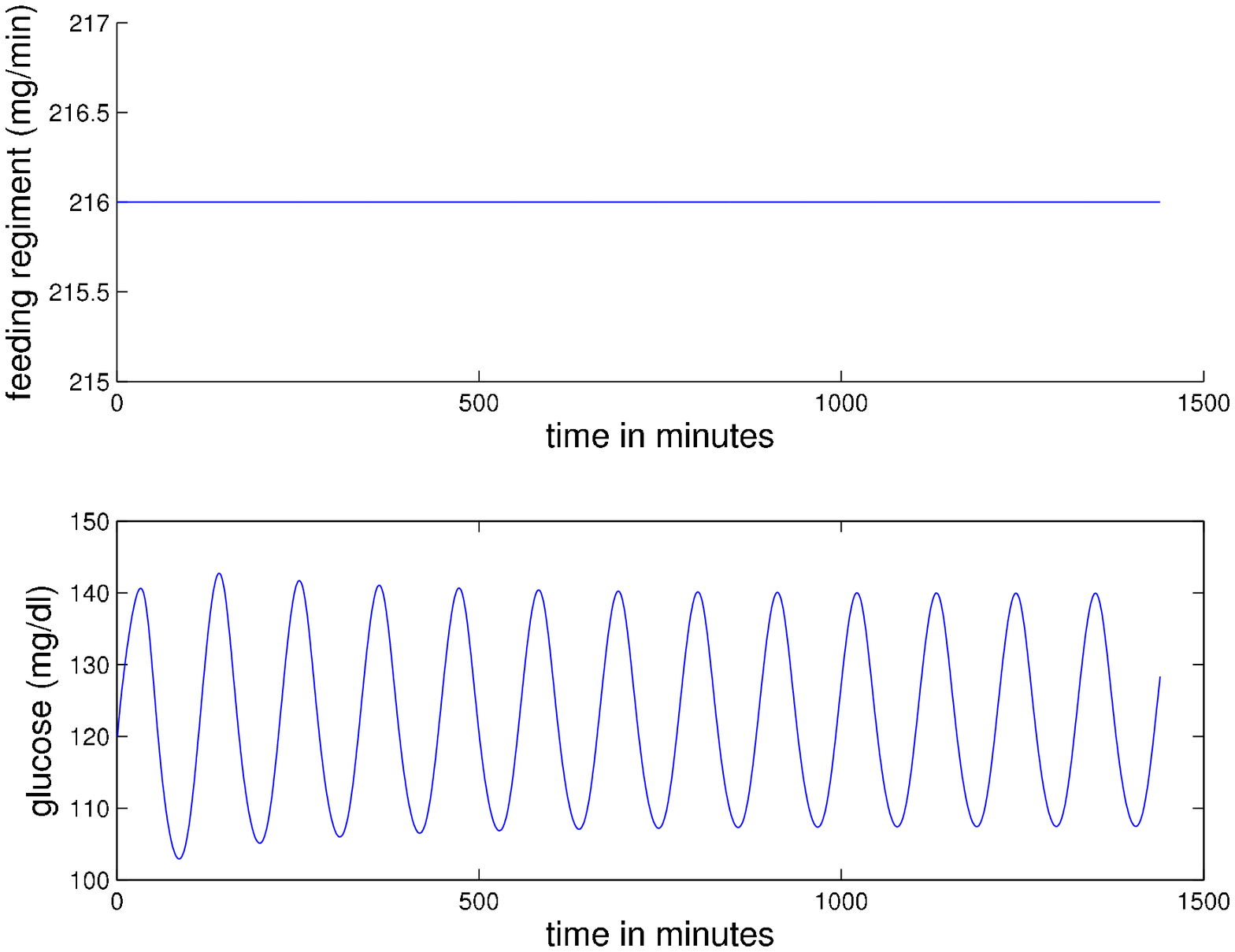, height=6cm}
    \label{fig:model_a}
  }
  \subfigure[Glucose-insulin model with three meals and glucose response]{
    \epsfig{file=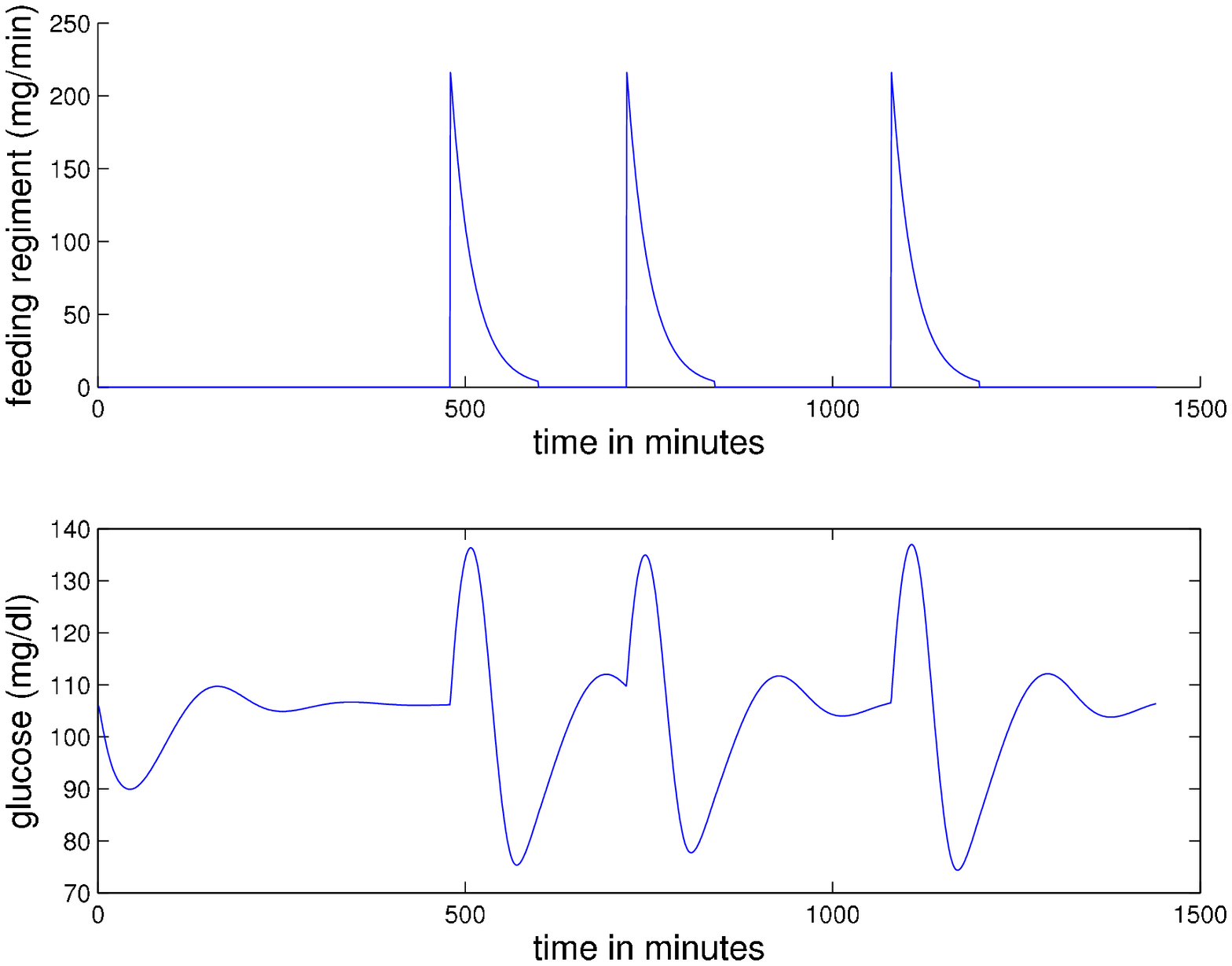, height=6cm}
    \label{fig:model_b}
  }  
  \caption{{\bf Depicted above are:} (a) the glucose for the standard
    glucose-insulin model with continuous feeding; and (b) the glucose for
    the standard glucose-insulin model with realistic meal structure.}
  \label{fig:model}
\end{figure}

Figure \ref{fig:model} details the feeding-glucose response for the
models.  The point of this figure is to depict the basic building
blocks that will be aggregated into a population. Figure
\ref{fig:model_a} demonstrates that, relative to the model, a
\emph{continuous} infusion of glucose induces a periodic oscillation
in intravascular glucose whose period is on the order of minutes; note
that verification of this signal in humans can be found in Fig. $1$ of
Ref. \cite{sturis_91} or more generally
Ref. \cite{glucose_osc_human_short_term_nejm}. Furthermore, note that
\emph{in this case} the glucose oscillation is \emph{exactly symmetric
  about its mean}, implying that long term averages of the
glucose-insulin response should be a constant --- this fits with the
intuitive control theory vision of the glucose-insulin cycle. Figure
\ref{fig:model_b} illustrates the glucose oscillation structure that
is induced when the feeding regiment consists of three realistic meals
given at $8$, $12$, and $18$ hundred hours respectively.  Note that
the peaks and length of time over which the glucose response exists
depends on the magnitude of the calories in the meal --- one way of
conceptualizing this system is as a forced oscillator with damping
that depends on caloric input and metabolism. Also note that the when
caloric intake is a pulse, the glucose-insulin response is \emph{not}
symmetric about the mean or baseline. In particular, isolating the
glucose response and integrating the response relative to the baseline
yields a very small but negative number, meaning that the overall
glucose level is depressed when integrated over the course of the meal
and response relative to this model.

\subsection{Diurnal variability of glucose in a population}

\begin{figure}[!ht]
  \subfigure[Normalized population glucose by hour.]{
    \epsfig{file=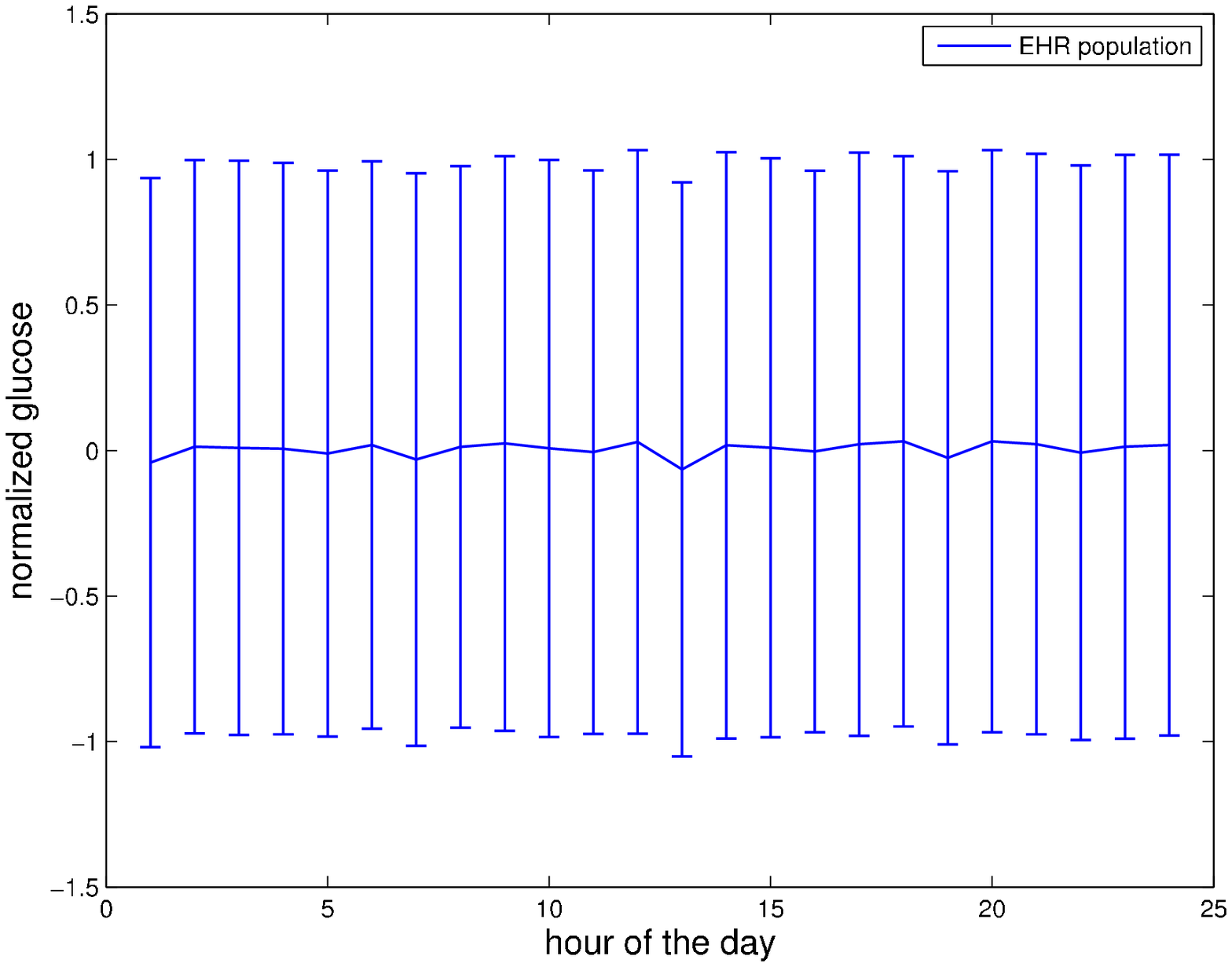, height=6cm}
    \label{fig:general_patients_a}
  }
  \subfigure[Single patient normalized glucose by hour.]{
    \epsfig{file=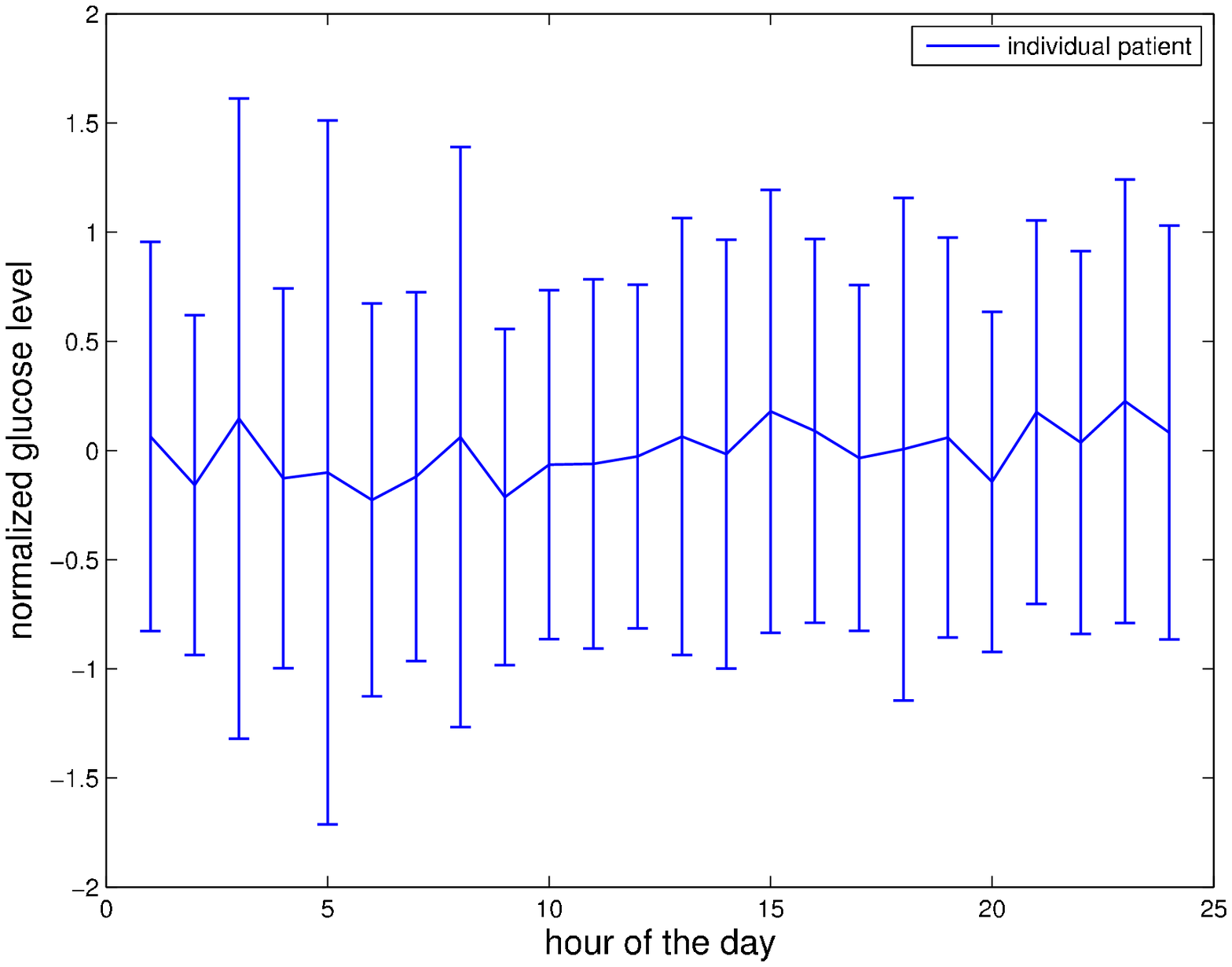, height=6cm}
    \label{fig:general_patients_b}
  }
  \subfigure[Normalized NICU population glucose and feeding by hour.]{
    \epsfig{file=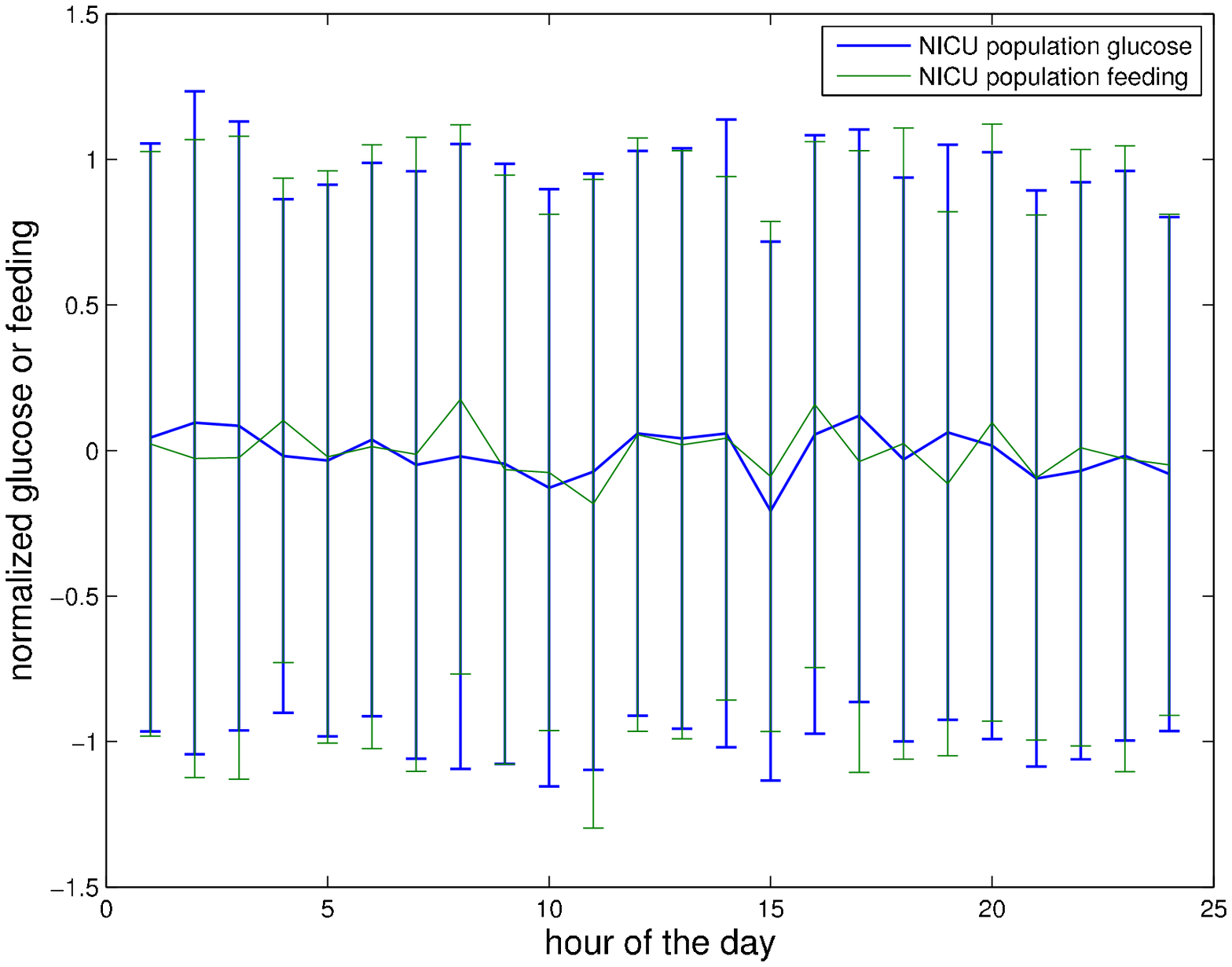, height=6cm}
    \label{fig:general_patients_c}
  }
  \subfigure[Normalized model glucose by hour]{
    \epsfig{file=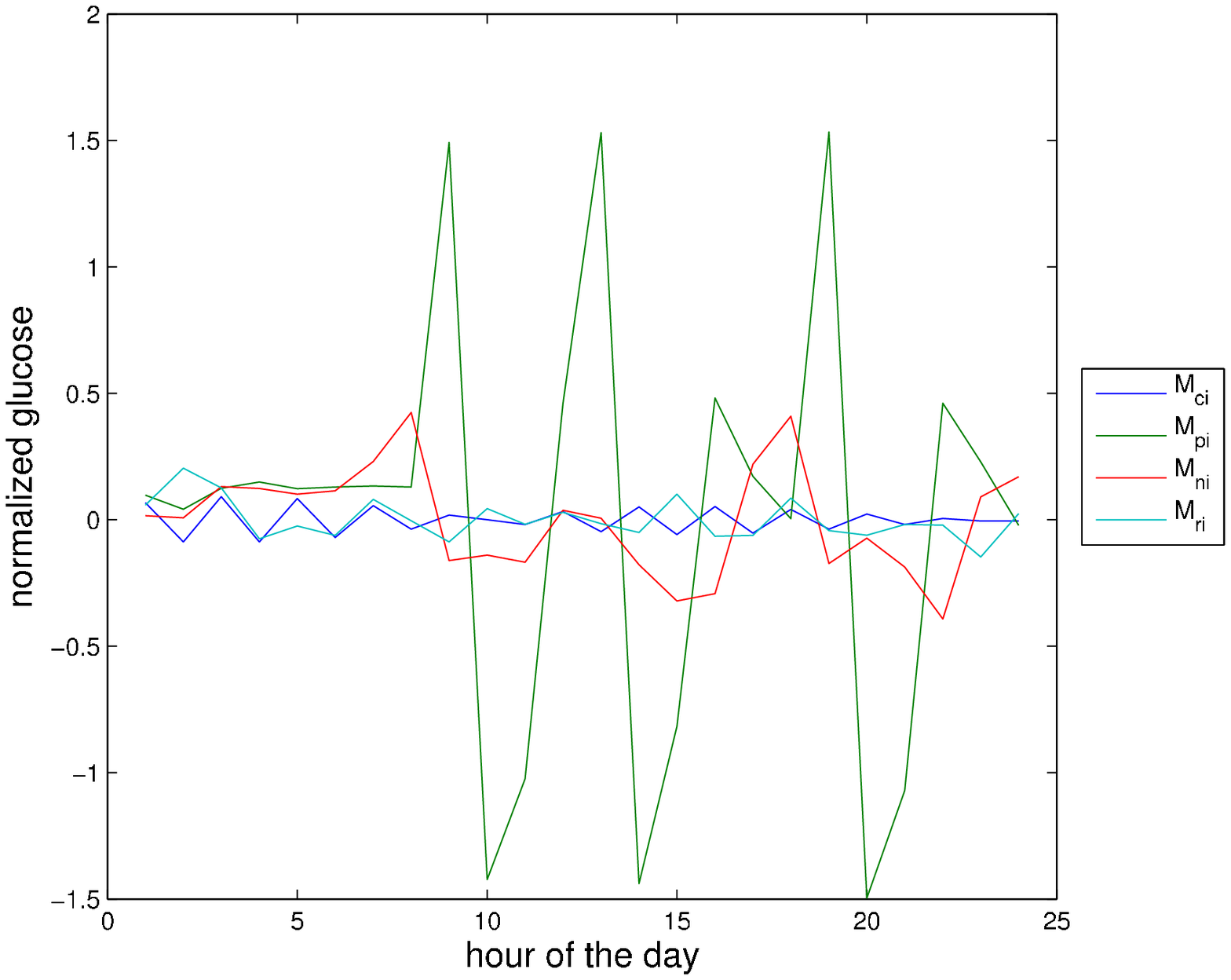, height=6cm}
    \label{fig:general_patients_d}
  }
  \caption{{\bf Depicted above are:} (a) the mean and standard deviation in
    glucose, by hour, for $800,000$ patients whom have been normalized
    to mean zero and variance one, with at least two glucose
    measurements from the CUMC EHR; (b) a single patient's --- the
    patient with the most glucose measurements in the CUMC EHR ---
    mean and standard deviation in glucose measurements by hour; (c)
    the mean and standard deviation in glucose and \emph{intravenous}
    feeding rates, by hour, for $43$ normalized patients in the neural
    ICU; (d) glucose, by hour, for various different model feeding
    regiments.}
  \label{fig:general_patients}
\end{figure}

With the basic building blocks of glucose-insulin response in place,
next consider Fig. \ref{fig:general_patients} which details the hourly
glucose variability within the data sets and models. In particular, in
Fig. \ref{fig:general_patients_a} the hourly glucose variability for
the EHR population displays \emph{no diurnal variability} or
signal. While we expected the short-term oscillations to average out
we also expected to observe a small but statistically significant
signal on a 24-hour cycle that matched meal times. More specifically,
we expected a small diurnal signal because: (i) humans eat
periodically, which, intuitively, implies that glucose would be
broadly higher over meal times; and (ii), there exists a \emph{weak
  but present diurnal variability} in kidney function that was
observed on the same data set
\cite{stat_dyn_diurnal_correlation_short} --- which was surprising in
and of itself because kidney function is not normally believed to have
a strong diurnal signal. The fact that there is no signal is due, at a
fundamental level, to the fact that the endocrine system always acts
as a fast controller; from the control theory perspective of the
glucose/insulin system, the lack of hourly glucose variability is
indeed what one might expect. Specifically, because the endocrine
system attempts to keep the glucose level constant, or within a given
interval, and because the endocrine system works on a fast time scale
(e.g., seconds to tens of minutes), the glucose/insulin
\emph{dynamics} of the endocrine system are relatively fast --- much
faster than an hour --- and thus small changes in meal times will
force glucose levels to average out when \emph{hourly} glucose levels
are averaged over many days.

A comparison of the data-based signals in
Figs. \ref{fig:general_patients_a} and \ref{fig:general_patients_b}
with the modeling results shown in Fig. \ref{fig:general_patients_d} leads
to the following observations/conclusions. \emph{First}, constant
feeding in the model for a population leads to constant (averaged by
hour) glucose which agrees with the data-based result (NICU patients)
of Fig.  \ref{fig:general_patients_b}, and thus verifies that relative
to hourly glucose variability, the model correctly predicts the
observations. \emph{Second}, the periodically driven individual has
the expected daily meal response structure; but the signal is too
clean to realistically represent an individual or a
population. \emph{Third}, the random feeding produces no diurnal
signal and thus agrees with the data-based result (wild population)
from Fig. \ref{fig:general_patients_a}, meaning that it is possible
either that the model does not depend on strongly on feeding structure
or that the by-hour glucose isn't good enough to detect feeding
structure and differentiate the respective populations. And
\emph{fourth} the noisy periodic case has wide, weak diurnal peaks at
meal times which differs from what is observed in the data; however,
the primary reason the diurnal structure in daily glucose variability
is retained in the models with noisy periodic-like feeding is that the
meals are uniformly distributed within two hour \emph{disjoint}
intervals. We know from further experiments that increasing the
\emph{diversity of the location of the mealtime windows between
  individuals, while retaining the noisy mealtime structure within
  individuals}, allows the model results to reproduce the population
signal shown in Fig. \ref{fig:general_patients_a} more faithfully.

Finally, considering the model output shown in Fig. \ref{fig:model_b}
where the glucose-insulin response to a meal \emph{is roughly
  symmetric about the baseline glucose level}, implying that the
\emph{average} glucose over many data points \emph{should} be at least
near zero when averaged over time and a population, the data-based
results in Fig. \ref{fig:general_patients} are not surprising. But,
\emph{what is a surprise} is that this simple model \emph{can
  accurately represent a population over much longer time periods than
  it was designed to represent.} Or, more specifically, while the
model we use here is an ultradian model designed to be applicable on a
time-scale of much less than a day, the model nevertheless appears to
be applicable over time periods considerably longer than a day.

\subsection{Diurnal variability in nonlinear correlation of glucose}

\begin{figure}[!ht]
  \subfigure[All data sets and models --- a global view of the TDMI]{
    \epsfig{file=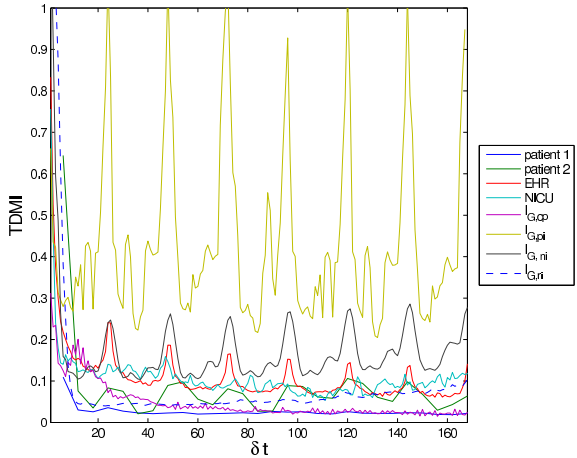, height=6cm}
    \label{fig:tdmi_a}
  }
  \subfigure[All data sets and models --- feeding scale TDMI for $\delta t$ of $1$ to $12$ hours]{
    \epsfig{file=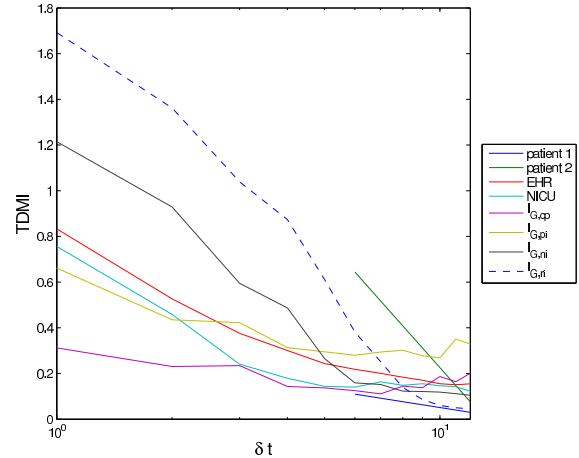, height=6cm}
    \label{fig:tdmi_b}
  }
  \subfigure[All data sets and models --- diurnal scale TDMI for
    $\delta t$ of $12$ to $72$ hours]{
    \epsfig{file=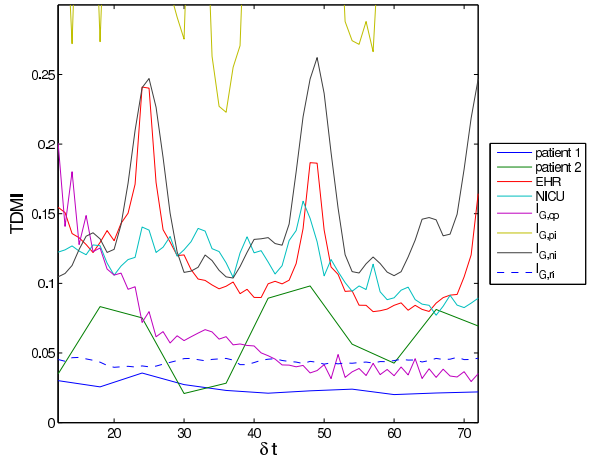, height=6cm}
    \label{fig:tdmi_c}
  }
  \caption{{\bf Depicted above are:} (a) the TDMI curves for all populations
    and models resolved to one hour intervals for time delays of up to
    one week, note the sharp decay in TDMI in all cases, and the
    diurnal peaks in all periodically fed populations or models; (b)
    the TDMI curves for all populations and models over time-delays of
    $1$ to $12$ hours; and (c) the TDMI curves for all populations and
    models from $12$ to $72$ hours, notice the diurnal peaks in all
    periodically fed populations or models.}
\label{fig:tdmi}
\end{figure}

Finally we arrive at the nonlinear-correlation variability in glucose
as quantified by the TDMI.  Figure \ref{fig:tdmi_a} frames the TDMI
over an entire seven day time-delay window. The following features of
\ref{fig:tdmi_a} are of note: \textbf{(i)} all models and data sets
show a sharp decay in TDMI between one and twelve hours; \textbf{(ii)}
one of the individual patients has weak diurnal peaks in the TDMI at
$24$ and $48$ hours while the other patient has diurnal peaks for
several days; \textbf{(iii)} the NICU population shows no long term
structure in the TDMI, although there does remain a constant amount of
TDMI present; \textbf{(iv)} the uncontrolled EHR population shows
diurnal peaks in the TDMI, and the magnitude of these peaks decays
with time; \textbf{(v)} the continuously fed population model, after
the decay within twelve hours, shows a weak hump at eighteen hours
that is a function of the exact symmetry of the periodic oscillations
in glucose, followed by a decay to small, constant, TDMI ---
\emph{thus, this model case accurately represents the NICU patients};
\textbf{(vi)} the periodic \emph{individual} model patient without
noise has a good deal of TDMI as well as sharp diurnal peaks and ---
note that from this it is self-evident that an individual patient with
a continuous feeding regiment would also have a high level of TDMI,
albeit without the sharp 24-hour peaks; \textbf{(vii)} noisy periodic
model has, after the sharp decay at twelve hours, diurnal peaks in the
TDMI with \emph{non-decaying} magnitude --- \emph{thus, this model
  mostly closely represents the real EHR population, and in fact the
  two overlay up to about $36$ hours}; \textbf{(viii)} the TDMI for
the randomly fed model case has no long term structure --- \emph{thus,
  the TDMI helps distinguish the constant feeding, the random feeding,
  and the noisy periodic feeding models.}  To consider more detailed
analysis, it is instructive to split Fig. \ref{fig:tdmi_a} into two
regimes, $\delta t < 12$ hrs, and $\delta t > 12$ hours.

The most important feature of Fig. \ref{fig:tdmi_b} is that the
collection of TDMI curves are \emph{bounded from above by the random
  feeding and below by the population with continuous feeding models}
respectively. The random meal case has the most TDMI within the first
$12$ hours because the random feeding case maximizes the amount of
observed TDMI per mealtime period. This maximization occurs for two
reasons: (i) isolated meals have a large amount of TDMI that persists
over approximately four hours; and (ii), meals are uniformly
distributed over the $24$ hour period and are unlikely to
overlap. Said simply, the TDMI for the random meals population with
$\delta t<12$ largely represents the pure \emph{intra-meal} TDMI,
which is the maximum TDMI amongst the models (and apparently real
populations) we examine. This argument is further backed-up by the
fact that the randomly fed population has the sharpest decay in
TDMI. The reason why the TDMI for the population of continuously fed
patient model is a lower-bound is due to a combination of aggregation
effects and superpositions of periodic orbits. To understand this,
recall Fig. \ref{fig:model_a} and note that each member of the
population of continuously fed patients will have orbits with
\emph{different amplitudes and frequencies} and that aggregating them
together at a given $\delta t$ will produce a distribution that will
closely resemble a uniform distribution --- the distribution that
minimizes TDMI over all distributions. All the other cases fit in
between these two extreme situations.

The longer the time (separation) scale is shown in
Fig. \ref{fig:tdmi_c} and includes the TDMI for all cases over
time-separations of $12$ to $72$ hours.  Begin by noting that there is
no structure in TDMI signal for the NICU population as well as the
random feeding and continuously fed population models.  Thus, using
only the TDMI and the normalized hourly glucose, it is difficult to
distinguish the continuously fed population from the randomly fed
population. In contrast, the EHR population, by displaying the diurnal
peaks, is easily distinguishable from the NICU population; thus the
TDMI helps distinguish the EHR and NICU populations in a way that
analysis of the raw glucose values could not.  Moreover, because the
noisy feeding and EHR populations strongly resemble one another (they
are nearly identical for $\delta t = [6, 36]$ hours), and because the
exactly periodic feeding yields far to much TDMI, \emph{the difference
  between the EHR population and the NICU population is likely due to
  noisy, but specifically structured (i.e., not totally random)
  meal times.} This conclusion both confirms that EHR data reproduces
what is believed to be the first order glucose dynamical effect,
nutrition, and verifies that the ultradian model can represent humans
for much longer time scales than hours to minutes. Finally, even the
individual patient with the weakest signal shows a peak at $24$ hours
and a weak peak at $48$ hours, which is consistent with the EHR-based
TDMI signal.

\subsection{Resultant synopsis}

Based on Fig. \ref{fig:tdmi_c}, \emph{the most basic and elemental
  result is thus: the model output can be used (in conjunction with
  the TDMI) to correctly predict the distinction between general EHR
  patients and NICU patients on time scales longer than a day.}
Moreover, the observed TDMI signal for the EHR population represents
\emph{noisy, but structured meal times over the population}; meaning,
\emph{we can detect human behavior patterns in EHR data and
  verify/test them with/against physiological models}. That is,
adjusting the feeding in the model alone was enough to account for the
difference in the observed TDMI signals and thus to distinguish the
populations, all without injecting difference (e.g., differences in
mean age) into the parameters. This implies that EHR data can at least
resolve some first order physiological effects. At a finer resolution,
while the first order moment of the TDMI (i.e., predictability) can be
used to separate the two populations of patients because of how
nutrition is ingested, understanding the second order moment is more
complicated and is beyond the scope of this paper. More explicitly, it
is likely that the higher order moments of the TDMI peaks will
depend, to some unknown level of detail, on the health state of the
patient. Moreover, because even narrow EHR populations are relatively
diverse and as yet unquantified in the context at hand, and because
even the simple model we used has $\sim 20$ parameters that we hold
fixed \emph{for all populations examined here} that are nevertheless
are available for variation, resolution of the higher order moments of
the TDMI peak is beyond the scope of the current paper. Nevertheless,
preliminary analysis seems to point to the TDMI being monotonically
dependent on nutrition and the functioning (or artificial regulation)
of the pancreas. Finally, we were able to use EHR data to test a
physiological model for a population, but, as is the case with many
other data-driven fields, derived values (i.e., the TDMI) were more
helpful than the raw values.

\section*{Discussion}

The end goal of population physiology is twofold: (i) we want to
derive population-scale, data-based signals over medium to long
time-scales in a way that can be connected to constructive,
mechanistic models to further the understanding of human physiology;
and (ii) we want to be able to use these verified, constructive,
mechanistic models to affect the health of human beings via clinical
care.  In this paper, we have demonstrated (i) but not (ii), primarily
because glucose/insulin modeling is not yet at a stage were it can be
applied to affect clinical care in a direct manner. Nevertheless, we
have begun one of the necessary steps for implementing (ii), we have
demonstrated that simple mechanistic models can accurately represent
humans over the longer time scales that are relevant to clinical
outcomes.

Scientifically, the results in this work demonstrate and imply that:
\textbf{(i)} the output from a simple glucose/insulin model can be
used to predict the difference between EHR and NICU patients over time
periods longer than a day; \textbf{(ii)} glucose measurements for a
population yield diurnal variation in \emph{correlation}, but glucose
measurements \emph{do not vary diurnally} when aggregated over time
and population because the glucose/insulin system behaves like a
control system whose fast time-scale dynamics occur on the order of
minutes; \textbf{(iii)} ``normal'' (here normal means, humans not
being fed intravenously) humans \emph{do} have a diurnal TDMI signal
in glucose; \textbf{(iv)} ``normal'' human glucose values \emph{do}
display an initial decay in correlations (between subsequent
measurements) to a relative baseline within $12$ hours; \textbf{(v)}
the models with the noisy but structured meal times match the
diurnal TDMI EHR signal, thus the diurnal cycle in predictability of
glucose is primarily driven by nutrition (not an internal clock);
\textbf{(vi)} EHR data can resolve a signal that spans multiple time
scales and can be used to test physiological models; \textbf{(vii)}
that the standard glucose/insulin model \cite{sturis_91} is applicable
beyond the time-spans it was designed for; \textbf{(viii)} the NICU
population and continuous feeding model TDMI signals match one another
--- in particular, humans being fed continuously \emph{do not} have a
diurnal TDMI signal or any structured signal at all; and \textbf{(ix)}
EHR data resolves human social behavior --- the structure of meal
times.

Looking forward, currently, population physiology suffers from the
lack of existent, time-dependent signals; discovering such signals
that can be related to physiological models is where many current
opens problems lie. Said differently, before one can go about refining
models and understanding dynamics mechanistically and over longer time
periods, one needs actual data-based signals, or stylized facts
\cite{kaldorstylizedfact61}, that can suggest and motivate refinements
in the models via \emph{testing} of those models. Nevertheless, this
does not negate the need for constructive modeling that allows for
either interaction with clinical care or better reflection of known
physiological problems --- for it is through qualitative understanding
of models as dynamical and control systems \cite{hybrid_ds} that
actionable clinical interventions will come. Relative to
glucose/insulin regulation, in some circumstances, monitoring and
correcting for hyperglyceimia can help reduce mortality significantly
\cite{vandenberghen_glucose_reg}. Nevertheless, correlation is not
causation; the \emph{mechanistic reasons} why glucose control in ICU
populations helps with outcomes is not well understood, and thus
optimal clinical interventions remain unavailable. The inevitable
conclusion is that glucose/insulin dynamics are poorly understood on
longer time scales; and moreover, the current state of glucose/insulin
physiological modeling does not have a mechanism for understanding the
fundamental physiological problems (i.e., longer term effects of
glucose dynamics) that can suggest productive clinical interventions
(e.g., ICU glucose control and regulation). But, again, such models
cannot be developed without impetus, and that impetus must come in the
form of concrete, data-based signals. While the data scarcity has made
such signals difficult to come by, EHR data will put the data scarcity
problems behind us and replace these problems with new signal
processing problems that must be overcome. This paper represents a
step forward in this direction by using EHR data to discover a
physiologic-based signal that is connected to physiologic-based models
even in the circumstance where direct observation of the physiological
variable does not yield a signal that can stratify the population.

\section*{Acknowledgments}
The authors would like to acknowledge the financial support provided
by NLM grant RO1 LM06910.



\end{document}